%%% Macros etc.
\magnification=\magstep1
\overfullrule=0pt
\def\q#1{\lbrack #1 \rbrack}
\def\pano{\par\noindent}
\def\smno{\smallskip\noindent}
\def\meno{\medskip\noindent}
\def\bigno{\bigskip\noindent}
\def\o#1{\overline{#1}}

\def\type{type$\,{\rm II}\ $}
\def\typea{type$\,{\rm II}\,{\rm A}\ $}
\def\typeb{type$\,{\rm II}\,{\rm B}\ $}

\def\cl{\centerline}

\def\section#1{\leftline{\bf #1}\vskip-7pt\line{\hrulefill}}
\def\bibitem#1{\parindent=8mm\item{\hbox to 6 mm{$\q{#1}$\hfill}}}
\def\P{\,\hbox{\hbox to -0.2pt{\vrule height 6.5pt width .2pt\hss}\rm P}}
\def\BNT{\,\hbox{\hbox to -0.2pt{\vrule height 6.5pt width .2pt\hss}\rm N}}
%%% real numbers
\def\BRT{\,\hbox{\hbox to -0.2pt{\vrule height 6.5pt width .2pt\hss}\rm R}}
%%% integer numbers
\def\BZT{{\rm Z{\hbox to 3pt{\hss\rm Z}}}}
\def\BZS{{\hbox{\sevenrm Z{\hbox to 2.3pt{\hss\sevenrm Z}}}}}
\def\BZSS{{\hbox{\fiverm Z{\hbox to 1.8pt{\hss\fiverm Z}}}}}

%%% rational numbers
\def\BQT{\,\hbox{\hbox to -2.8pt{\vrule height 6.5pt width .2pt\hss}\rm Q}}
\def\BQS{\,\hbox{\hbox to -2.1pt{\vrule height 4.5pt width .2pt\hss}$
 \scriptstyle\rm Q$}}
\def\BQSS{\,\hbox{\hbox to -1.8pt{\vrule height 3pt width
 .2pt\hss}$\scriptscriptstyle \rm Q$}}

%%% complex numbers
\def\BCT{\,\hbox{\hbox to -3pt{\vrule height 6.5pt width .2pt\hss}\rm C}}
\def\BCS{\,\hbox{\hbox to -2.2pt{\vrule height 4.5pt width .2pt\hss}$ 
 \scriptstyle\rm C$}}
\def\BCSS{\,\hbox{\hbox to -2pt{\vrule height 3.3pt width
 .2pt\hss}$\scriptscriptstyle \rm C$}}

%%% Commands for drawing
\def\putbox#1#2#3{\setbox117=\hbox{#3}
 \dimen121=#1mm \dimen122=#2mm \dimen123=\wd117 \dimen124=\ht117
 \divide\dimen123 by -2  \divide\dimen124 by -2
 \advance\dimen121 by \dimen123  \advance\dimen122 by \dimen124
 \rlap{\kern\dimen121\raise\dimen122\hbox{#3}}}
%%% Authors
\def\alda{1}
\def\antone{2}
\def\antonz{3}
\def\aspina{4}
\def\aspinb{5}
\def\selfa{6}
\def\selfb{7}
\def\diska{8}
\def\ferrara{9}
\def\gepe{10}
\def\greene{11}
\def\intri{12}
\def\kachru{13}
\def\kakl{14}
\def\kaplu{15}
\def\klemm{16}
\def\lyn{17}
\def\sche{18}
\def\schz{19}
\def\schd{20}
\def\schv{21}
\def\sei{22}
\def\stro{23}
\def\vafa{24}
%%% Fonts for title page

\font\Large=cmr12 scaled \magstep3
%%% Title Page
\rm
\nopagenumbers
\pano
{\rightline {\vbox{\hbox{hep-th/9601050}
                   \hbox{BONN-TH-96-01}
                   \hbox{IFP-606-UNC}
                   \hbox{January 1996}}}}
\bigno\bigno
\centerline{\Large Exactly Solvable Points in the Moduli}
\vskip 10pt 
\centerline{\Large Space of Heterotic N=2 Strings}
\vskip 1.0cm
\centerline{{Ralph Blumenhagen${}^1$}\ \  and \ \
            {Andreas Wi{\ss}kirchen${}^2$}}
\vskip 1.0cm
\centerline{${}^1$ \it Institute of Field Physics, Department of Physics
and Astronomy,}
\centerline{\it University of North Carolina,
Chapel Hill NC 27599-3255, USA}
\vskip 0.1cm
\centerline{${}^2$ \it Physikalisches Institut der Universit\"at Bonn,
Nu{\ss}allee 12, 53115 Bonn, Germany}
\vskip 1.0cm
\centerline{\bf Abstract}
\meno
We investigate the subset of exactly solvable $(0,4)$ world sheet 
supersymmetric string vacua contained in a recent class of Gepner-like 
$(0,2)$ superconformal models. The identification of these models with 
certain points of enhanced gauge symmetry on $K_3\times T_2$ can be 
achieved completely. Furthermore, we extend the construction of in 
general $(0,2)$ supersymmetric exactly solvable models to the case where 
also a nontrivial part of the vector bundle is embedded into the hidden 
$E_8$ gauge group. For some examples we explicitly calculate the enhanced 
gauge symmetries and show that they open up the way to interesting 
branches of the $N=2$ moduli space. For some of these models candidates of
\type dual descriptions exist. 
\footnote{}
{\pano
${}^1$ e-mail:\ blumenha@physics.unc.edu
\pano
${}^2$ e-mail:\ wisskirc@avzw02.physik.uni-bonn.de}
\vfill\eject
%%% text
\footline{\hss\tenrm\folio\hss}
\pageno=1
\pano
\section{1.\ Introduction}
\meno
In two recent papers [\selfa,\selfb] we have presented a class of exactly 
solvable heterotic string models with in general $(0,2)$ supersymmetry 
for the internal superconformal field theory (SCFT). The method used to 
construct these purely heterotic modular invariant partition functions was
based on the simple current technique developed by Schellekens and 
Yankielowicz [\sche,\schz,\schd,\schv] and independently by Intriligator 
[\intri]. Since the input data of a SCFT are completely different from the
ones used in the (non)linear $\sigma$-model approach for strings in 
background fields, it turned out to be quite a difficult task to identify 
models from both sets. However, for at least three models such a one to one
correspondence could be achieved by comparing the chiral rings of these 
$N=1$ space-time supersymmetric string vacua [\selfb]. Therefore, we 
strongly believe that the class of $(0,2)$ $\sigma$-models does really 
provide us with fully fledged classical vacua of the string disproving, at 
least for certain points in the moduli space, earlier indications that 
they are destabilized by world sheet instanton effects.
\pano 
During the last months heterotic models with $N=2$ space-time 
supersymmetry have received wide attention due to their relevance for a 
stringy generalization of the Seiberg/Wit\-ten duality for $N=2$ 
Yang-Mills theories [\sei]. There exist convincing arguments that some 
heterotic strings on $K_3\times T_2$ admit a dual description in terms of
a \typea string compactified on a Calabi-Yau threefold 
[\antone,\antonz,\ferrara,\kachru,\kakl,\kaplu,\vafa] with the special 
property of being a $K_3$ fibration [\aspina,\klemm]. Nonperturbative 
topology changing transitions between different Calabi-Yau moduli spaces 
[\aspinb,\greene,\stro] for the \type string correspond on the heterotic 
side to the existence of points of enhanced gauge symmetry at which new 
Higgs phases appear. 
\pano
We will address several questions in this letter. First, we will construct
exactly solvable SCFTs with $N=2$ space-time supersymmetry. Since there is 
only one suitable manifold for the compactification, an identification of 
our models can be carried out much easier compared to the $N=1$ case, for 
one only has to find the vector bundle on $K_3$ for the left moving 
$\sigma$-model fermions. Furthermore, from earlier calculations it is 
known that exactly solvable SCFTs usually correspond to points in the 
moduli space where enhanced gauge symmetries occur. For instance, the usual
Gepner models [\gepe] have at least a further $U(1)^{r-1}$ piece in the 
gauge group. We will construct and analyze SCFTs, for which also 
nonabelian groups occur. By the Higgs mechanism we can break parts of the 
enhanced gauge symmetry and arrive at models for which a dual \type string
appears in the lists of $K_3$ fibrations in [\klemm,\lyn]. A similar 
calculation has been carried out in [\alda] using orbifold techniques 
for $T_4/Z_M \times T_2$. Thus, our results can be considered as the 
generalization of their analysis to Gepner-like SCFTs. 
\pano
It is clear that as long as the hidden gauge group $E_8$ is not broken 
one can only achieve a very limited set of models, for the number of 
generic vector multiplets is forced to be larger than nine. Thus, we will 
further address the problem, how the embedding of a vector bundle into the 
hidden $E_8$ factor can be implemented into the modular invariant partition
functions constructed in [\selfa]. An identification of these models with 
certain bundles on $K_3$ is straightforward and the Higgs phase of such 
models provides us with a lot of \type duals. 
\vfill\eject\pano
\section{2.\ Exactly solvable N=2 string models}
\meno
It is well known that in order to get $N=2$ space-time supersymmetry in 
the heterotic string compactified to four dimensions one needs a toroidal 
$c=3$ piece and at least a $(0,4)$ world sheet supersymmetric $c_r=6$ piece
which corresponds to the $K_3$ manifold. The left moving central charge 
$c_l$, however, can vary reflecting the freedom of choosing vector bundles 
of arbitrary rank for the left moving $\sigma$-model fermions. 
\pano
In this section we will first calculate a bunch of examples of exactly 
solvable $(0,4)$ models with generic $E_7\times E_8$ gauge group and by 
comparing the massless spectra we will try to identify them with suitable
bundles on $K_3$. To this end, we review how the massless spectrum is 
determined for general bundles on $K_3$.
\pano
The unbroken gauge group in ten dimensions should be $G=E_8\times E_8$. 
Now, one embeds a (complexified) stable, holomorphic vector bundle $V$ 
into ${\cal G}$, which breaks it down to a subgroup $G$ such that
$G\times H\subset{\cal G}$ is the maximal subalgebra. Then one can 
calculate the massless matter spectrum as follows:\ decompose
$$ {\rm adj}({\cal G})=\sum_i (M_i,R_i) \eqno(2.1) $$ 
where $M_i$ and $R_i$ are representations of $G$ and $H$, respectively. 
Then, the number of left-handed spinor multiplets transforming in the 
$M_i$ representation of $G$ is given by the Atiyah-Singer index theorem:
$$ N_{M_i}={\rm dim}(R_i)-{1\over 2} \int_{K_3} c_2(V)\,\,\,
 {\rm index}(R_i) .\eqno(2.2) $$
Furthermore, there appear 20 gauge singlet hypermultiplet moduli arising 
from higher dimensional gravitational fields. For the $SU(2)$ tangent 
bundle on $K_3$ one has $\int_{K_3} c_2(V)=24$. If one first 
compactifies the ten-dimensional theory on a torus, there exists the 
possibility of embedding vector bundles $V_a$ also into the possibly 
extended gauge group factors of the torus, $SU(3)_T$, $SU(2)^2_T$ and 
$SU(2)_T$. Then the following conditions have to be satisfied:
$$ c_1(V_a)=0,\quad \sum_a c_2(V_a)=c_2(TK_3)=24 \eqno(2.3)$$
so that the bundles admit spinors and satisfy the Donaldson-Uhlenbeck-Yau 
equation and the anomaly cancellation condition, respectively. Using our 
stochastic search computer program for the construction of exactly 
solvable models with $N=2$ space-time supersymmetry we find a lot of 
models with all kinds of gauge groups. For instance, we obtain a model 
with gauge group $SO(10)\times E_8$ where the vacuum sector contains 
$N_g=15$ further massless vectors\footnote{$^1$}{Actually, there are 14 
vector multiplets including the dilaton multiplet plus the 
gra\-vi\-pho\-ton from the graviton multiplet. The graviphoton and the 
vector in the dilaton multiplet form a $U(1)^2_T$ part of the gauge 
group.}, which also form a to be determined gauge group 
$G(13)\times U(1)^2_T$ of dimension $15$. Furthermore, the massless 
sector contains $N_{16}=10$ hypermultiplets in the spinor 
representation, $N_{10}=12$ hypermultiplets in the vector representation
and $N_1=100$ hypermultiplets in the singlet representation of $SO(10)$.
On the one hand side, by embedding an $SU(4)$ bundle with 
$A^{E_8}_4=\int_{K_3} c_2(V)=18$ into $E_8$ and an $SU(3)$ 
bundle with $A^{SU(3)}_3=6$ into the enhanced toroidal $SU(3)_T$ gauge 
group one gets the gauge group $G'=SO(10)\times E_8\times U(1)^2_T$ with 
the following massless spectrum:\ $N_{16}=10$, $N_{10}=12$, $N_1=87$. By 
giving vacuum expectation values (VEVs) to exactly $r=13$ of our 
hypermultiplet moduli we can break the enhanced gauge group of dimension
$15$ down to $U(1)_T^2$ and thus reaching complete agreement of the two 
spectra. On the other hand side, this model can be understood as lying
in the following pattern of successive symmetry breakings: 
%%% begin of figure
$$
  \putbox{0}{0}{$E_7$}
  \putbox{8}{0}{$\longrightarrow$}
  \putbox{16}{0}{$E_6$}
  \putbox{24}{-1}{$\searrow$}
  \putbox{32}{-5}{$SO(10)$}
  \putbox{40}{-1.5}{$\nearrow$}
  \putbox{48}{0}{$SU(5)$}
  \putbox{56}{-1}{$\searrow$}
  \putbox{64}{-1.5}{$SU(4)$}
  \putbox{64}{-5.5}{$\cong$}
  \putbox{72}{-1.5}{$\nearrow$}
  \putbox{80}{0}{$SU(3)$}
  \putbox{88}{0}{$\longrightarrow$}
  \putbox{96}{0}{$SU(2)$}
  \putbox{104}{-1}{$\searrow$}
  \putbox{108}{-5}{$0$}
  \putbox{123}{-5}{$(2.4)$}
  \putbox{-0.5}{-10}{$SO(14)$}
  \putbox{8}{-10}{$\longrightarrow$}
  \putbox{16.5}{-10}{$SO(12)$}
  \putbox{24}{-8.5}{$\nearrow$}
  \putbox{40}{-8}{$\searrow$}
  \putbox{48}{-10}{$SO(8)$}
  \putbox{56}{-8.5}{$\nearrow$}
  \putbox{64}{-8.5}{$SO(6)$}
  \putbox{72}{-8}{$\searrow$}
  \putbox{80}{-10}{$SO(4)$}
  \putbox{88}{-10}{$\longrightarrow$}
  \putbox{96}{-10}{$SO(2)$}
  \putbox{104}{-8.5}{$\nearrow$}
  \hskip 11.5cm \hfill
$$
%%% end of figure 
The corresponding $E_7$ model has $N_{56}=7$ and $N_{1}=76$ 
hypermultiplets and also a further gauge group of dimension $N_g=15$. 
\pano
We find another example with gauge group $E_8\times E_8\times
U(1)^2_T\times G(10)$ and $N_1=84$ hypermultiplets in the singlet 
representation. This model can be understood by embedding an $SU(3)$
bundle with $A_3^{SU(3)}=24$ into the toroidal $SU(3)_T$. 
\pano
In Table 2.1 we enumerate the interpretation in terms of bundles on 
$K_3$ for some of the exactly solvable models we obtained. We always 
list only the $E_7$ parent model and indicate in column six at which 
special level in the cascade of symmetry breaking our exactly solvable
model occurs. 
\bigno
\cl{\vbox{ 
\hbox{\vbox{\offinterlineskip
\def\tablespace{height2pt&
 \omit&&\omit&&\omit&&\omit&&\omit&&\omit&&\omit&\cr}
\def\tablerule{\tablespace\noalign{\hrule}\tablespace}
\hrule\halign{&\vrule#&\strut\hskip0.1cm\hfil#\hfill\hskip0.05cm\cr
\tablespace
& $N_{56}$ && $N_1$ && $N_g$ && bundle && enhanced gauge group && $G'$ && 
$r$ &\cr
\tablerule\tablerule
& $10$ && $65+r$ && $10+r$ && $A_2^{E_8}=24$ && 
 $G'\times SU(3)_T\times U(1)^2_T\times G(r)$ && $E_7, E_6, SO(12)$ && 
 $9,6,4,2$ & \cr
\tablerule
& $8$ && $62+r$ && $3+r$ && $A_2^{E_8}=20$ && 
 $G'\times U(1)^3_T\times G(r)$ && $SO(12), SO(10)$ && $19,14,10$ & \cr
& && && && $A_2^{SU(2)}=4$ && && && $7,5$ &\cr
\tablerule
& $7$ && $63+r$ && $2+r$ && $A_2^{E_8}=18$ && 
 $G'\times U(1)^2_T\times G(r)$ && $E_6, SO(10)$ && $13,8,5$ & \cr
& && && && $A_3^{SU(3)}=6$ && && && $$ &\cr
\tablerule
& $6$ && $65+r$ && $2+r$ && $A_2^{E_8}=16$ && 
 $G'\times U(1)^2_T\times G(r)$ && $SO(12) $ && $14$ & \cr
& && && && $A_3^{SU(3)}=8$ && && && $$ &\cr
\tablerule
& $4$ && $69+r$ && $2+r$ && $A_2^{E_8}=12$ && 
 $G'\times U(1)^2_T\times G(r)$ && $E_7, E_6$ && $17,13,6$ & \cr
& && && && $A_3^{SU(3)}=12$ && && && $14,12$ &\cr
\tablerule
& $2$ && $59+r$ && $2+r$ && $A_2^{E_8}=8$ && 
 $G'\times U(1)^2_T\times G(r)$ && $SO(14)$ && $21$ & \cr
& && && && $A_{2,2}^{SU(2)^2}=16$ && && && $$ &\cr
\tablerule
& $1$ && $75+r$ && $2+r$ && $A_2^{E_8}=6$ && 
 $G'\times U(1)^2_T\times G(r)$ && $E_7$ && $7$ & \cr
& && && && $A_3^{SU(3)}=18$ && && && $$ &\cr
\tablespace}\hrule}}
\hbox{\hskip 0.5cm Table 2.1 \hskip 0.5cm Exactly solvable $N=2$ spectra}}}
\meno
Note that the appearance of the exceptional $SU(2)_T\times U(1)_T$ gauge 
group for the torus in the $N_{56}=8$ example is not obvious at all from 
the numbers in Table 2.1, for they would also fit with an $SU(3)_T$
gauge group. However, as we will see in the following section, the massless
spectrum does indeed contain a further $U(1)$ current, relative to which 
all other fields are uncharged. Certainly, our list of $(0,4)$ models 
obtainable from the $(0,2)$ generalized Gepner models is not complete. 
However, it already shows that probably all models would fit into the 
scheme of $K_3$ compactifications. Thus we are confident that this nice 
picture will also hold for all $N=1$ supersymmetric SCFTs, even though a 
case by case identification with bundles on arbitrary Calabi-Yau 
threefolds is hardly to reveal. 
\bigno
\section{3.\ Special points of enhanced gauge symmetry}
\meno
After identifying exactly solvable $(0,4)$ string models with certain 
bundles on $K_3\times T_2$ we draw our attention to the complete gauge 
symmetry of these models. In the Gepner case this gauge group is in most 
cases only a further $U(1)^{r-1}$ abelian group. However, we will see that 
this is not necessarily the case if one has introduced further simple 
currents in the modular invariant partition function. The motivation is, 
that points of enhanced symmetry turned out to be an important issue in 
the study of how the moduli space of heterotic $N=2$ strings is connected.
By giving VEVs to various kinds of scalar fields both in the 
hypermultiplets and the vector multiplets one can move from one branch of 
the moduli space to another one thereby changing the spectrum $(n_H,n_V)$
drastically. Thus, in order to gain information about the points of 
enhanced symmetry on $K_3$, one can use SCFTs as a natural source for 
such points.
\pano
In [\kachru] it was argued that some heterotic $N=2$ models have a dual 
description as \typea models compactified on a Calabi-Yau manifold. Since 
in the \typea string the dilaton is a member of a hypermultiplet, the 
spectrum is given by $(n_H,n_V)=(b_{21}+1,b_{11}+1)$. If such a duality 
holds one can calculate the full quantum moduli space of the 
vector multiplets on the \typea side and the quantum moduli space of the
hypermultiplets on the heterotic side. Thus, analogously to mirror 
symmetry one has a handle on the complete quantum corrected field theory 
by using this complementary picture.
\bigno
{\bf 1.\ Example based on the $(1)^3\,(4)^3$ parent}
\meno
In the following we will study in detail some of the models in the lists 
above. We start with the model with gauge symmetry 
$SO(10)\times E_8\times U(1)_T^2\times G(13)$ and the massless spectrum 
$N_{16}=10$, $N_{10}=12$ and $N_1=100$. Inspection of the couplings of the 
right moving vacuum in the modular invariant partition function shows that 
besides the 6 $U(1)$ currents from each of the unitary factors there are
exactly seven further currents around. All these together can be shown to 
form an $SU(3)_E\times U(1)_E^5$ algebra, so that the fully enhanced gauge 
symmetry at this point of the moduli space is 
$$SO(10)\times E_8\times SU(3)_E\times U(1)^5_E\times U(1)_T^2.\eqno(3.1)$$
After we have determined the complete enhanced gauge symmetry of the model 
we have to show how the massless sector fits into irreducible 
representations of $SO(10)\times SU(3)_E$. The answer is as follows:
$$(16,3)+7\times(16,1)+(10,3)+9\times(10,1)+28\times(1,3)+16\times(1,1).
 \eqno(3.2)$$
We have suppressed all the $U(1)^5_E$ charges, where for every $U(1)_E$ 
factor there exists at least one state carrying nonzero charge under this 
current. Thus, they can be higgsed away completely. 
\pano
The information in (3.2) allows us to investigate the Higgs sector of 
the model by successively giving VEVs to some fields in the fundamental 
representation of the gauge group such that it is broken down to some 
subgroup. Then, by giving VEVs to the adjoint representation of the 
remaining gauge group we move to a generic point in the vector moduli and 
the massless spectrum is given by two numbers, the number of 
hypermultiplets and the number of vector multiplets. In Table 3.1 we 
collect all models which occur in the Higgs phase of this model and have 
an appropriate \typea dual in the lists of [\klemm,\lyn]. These are 
approximately $6\%$ of all spectra occurring in the entire chain.
The second column features the remaining enhanced gauge group of the model
before moving to a generic point in the vector moduli.
\bigno
\cl{\vbox{
\hbox{\vbox{\offinterlineskip
\def\tablespace{height2pt&\omit&&\omit&&\omit&\cr}
\def\tablerule{\tablespace\noalign{\hrule}\tablespace}
\hrule\halign{&\vrule#&\strut\hskip0.2cm\hfil#\hfill\hskip0.2cm\cr
\tablespace
& $(n_H,n_V)$ && 
 $G'\subset SO(10)\times E_8\times SU(3)_E\times U(1)_E^5\times U(1)^2_T$ 
 && $K_3$ fibration &\cr
\tablerule\tablerule
& $(322,10)$ && $E_8\times U(1)_T^2$ && $\P_{1,1,6,16,24}[48]$ & \cr
\tablespace
& $(221,11)$ && $SU(2)\times E_8\times U(1)_T^2$ && 
 $\P_{1,1,6,10,18}[36]$ & \cr
\tablespace
& $(168,12)$ && $SU(3)\times E_8\times U(1)_T^2$ && 
 $\P_{1,1,6,10,12}[30]$ & \cr
\tablespace
& $(129,13)$ && $SU(4)\times E_8\times U(1)_T^2$ && 
 $\P_{1,1,6,10,12,14}[24\,20]$ & \cr
\tablerule
& $(144,12)$ && $SU(2)\times E_8\times SU(2)_E\times U(1)_T^2$ && 
 $\P_{1,1,6,8,10}[26]$ & \cr
\tablerule
& $(55,13)$ && $SO(2)\times E_8\times SU(3)_E\times U(1)_T^2$ && 
 $\P_{2,2,5,5,6}[20]$ & \cr
\tablespace
& $(50,14)$ && $SO(4)\times E_8\times SU(3)_E\times U(1)_T^2$ && 
 $\P_{5,5,8,12,30}[60]$ & \cr
\tablespace
& $(33,21)$ && $SO(8)\times E_8\times SU(3)_E\times U(1)_E^5
 \times U(1)_T^2$ && $\P_{3,3,4,4,10}[24]$&\cr
\tablespace}\hrule}}
\hbox{\hskip 0.5cm Table 3.1 \hskip 0.5cm Heterotic/\typea duals}}}
\meno
The cascade of the first four models has already been found in [\alda]. 
However, the second four models make use of the special enhanced gauge 
group $SU(3)_E\times U(1)_E^5$ and therefore lead to new branches of the 
$N=2$ moduli space. 
\bigno
{\bf 2.\ Example based on the $(1)\, (4)^4$ parent}
\meno
As a second example we study the model with gauge group 
$SO(10)\times E_8\times U(1)^2_T\times G(8)$ and the massless spectrum 
$N_{16}=12$, $N_{10}=14$ and $N_1=97$. By looking at the couplings to the 
right moving vacuum state, one realizes that there appears an enhanced 
$SU(2)_E\times U(1)$ gauge group, where the entire massless spectrum 
contains no state charged under this $U(1)$. This indicates that this model
is actually the one where the torus gauge group is $SU(2)_T\times U(1)^3_T$
with a bundle with $A^{SU(2)}_2=4$ embedded into the $SU(2)_T$ part. The 
massless spectrum fits into the following representations of 
$SO(10)\times SU(2)_E$:
$$\eqalignno{
 &4\times(16,2)+4\times(16,1)+(10,3)+4\times(10,2)+3\times(10,1)+&\cr
 &4\times(1,4)+11\times(1,3)+16\times(1,2)+16\times(1,1)&(3.3)\cr}$$
which is somehow surprising for the appearance of higher spin $SU(2)$ 
representations. Thus, appropriate simple currents give rise to the 
existence of higher level Kac-Moody algebras in the spectrum, in this case 
$SU(2)_3$. 
\pano
As in the previous example, we want to see what happens in the different 
Higgs phases of this model which are summarized in Table 3.2. 
\bigno
\cl{\vbox{
\hbox{\vbox{\offinterlineskip
\def\tablespace{height2pt&\omit&&\omit&&\omit&\cr}
\def\tablerule{\tablespace\noalign{\hrule}\tablespace}
\hrule\halign{&\vrule#&\strut\hskip0.2cm\hfil#\hfill\hskip0.2cm\cr
\tablespace
& $(n_H,n_V)$ && 
 $G'\subset SO(10)\times E_8\times SU(2)_E\times U(1)_E^4 \times U(1)^3_T$ 
 && $K_3$ fibration &\cr
\tablerule\tablerule
& $(377,11)$ && $E_8\times U(1)_T^3$ && $\P_{1,1,8,20,30}[60]$ & \cr
\tablespace
& $(252,12)$ && $SU(2)\times E_8\times U(1)_T^3$ && 
 $\P_{1,1,8,12,22}[44]$ & \cr
\tablespace
& $(187,13)$ && $SU(3)\times E_8\times U(1)_T^3$ && 
 $\P_{1,1,8,12,14}[36]$ & \cr
\tablespace
& $(140,14)$ && $SU(4)\times E_8\times U(1)_T^3$ && 
 $\P_{1,1,8,12,14,16}[28\, 24]$ & \cr
\tablespace
& $(117,15)$ && $SO(8)\times E_8\times U(1)_T^3$ && 
 $\P_{3,3,8,28,42}[84]$ & \cr
\tablespace}\hrule}}
\hbox{\hskip 0.5cm Table 3.2 \hskip 0.5cm Heterotic/\typea duals }}}
\meno
The first four models were also found in [\alda].
\bigno
{\bf 3.\ Example based on the $(1)^3\,(4)^3$ parent}
\meno
We briefly discuss a third example with gauge group $SO(12)\times E_8
\times SU(3)_E\times SU(2)_E\times U(1)^3_E\times U(1)^2_T$
at the exactly solvable point. The massless sector fits into the following 
representations of $SO(12)\times SU(3)_E\times SU(2)_E$:
$$\eqalignno{&(32,1,2)+2\times(32,1,1)+2\times(12,3,1)+2\times(12,1,2)+
 2\times(12,1,1)+&\cr &4\times(1,3,2)+10\times(1,3,1)+4\times(1,1,2)+
 14\times(1,1,1).&(3.4)\cr}$$
The only possible candidate for a \typea dual model is reached by first 
higgsing the $SU(2)_E\times U(1)^3_E$ away, then move to generic point in 
the $SU(3)_E$ vector moduli and break the $SO(12)$ down to $SU(2)$. This 
yields at generic points in the $SU(2)$ moduli the spectrum 
$(n_H,n_V)=(97,13)$ which could be dual to the \typea string compactified
on the $K_3$ fibration $\P_{1,1,6,8,8,10}[18\,16]$.
\bigno
Summarizing, by studying different branches of the $N=2$ moduli space of 
these 3 special models we already found several possible candidates 
for heterotic duals of \typea strings. One apparent limitation is that so 
far we have not broken the hidden $E_8$ symmetry in the exactly solvable 
approach, thus $n_V$ is always larger than nine. To overcome this, we have 
to extend our construction scheme for $(0,2)$ string models right from
the beginning.
\bigno
\section{4.\ Breaking the hidden $E_8$ }
\meno
Geometrically it is clear that one can get much more general gauge groups 
$G'$ by embedding a vector bundle into the so far hidden $E_8$ factor, as 
well. Since this implies that also non-singlet representations of $G'$ are
allowed to occur in the massless spectrum, it is possible to reduce the 
number of vector multiplets considerably by higgsing away parts of the 
broken gauge group $G'$. The question is, whether we can extend the 
formalism of $(0,2)$ exactly solvable string models proposed in [\selfa] to
this general case. 
\pano
In the spirit of the construction, we describe the embedding of an
$SU(2)$ bundle into $E_8$ by a spectral flow extension of 
$SO(12)$ to $E_7$. Surely, the spectral flow must have the right values 
$(H,Q)=({1\over4},1)$ of conformal dimension and $U(1)$ charge. This can 
be achieved by introducing two $U(1)_2$ factors which together with the 
$SO(12)$ part add up to the correct central charge $c=8$ of the $E_8$ 
current algebra. Thus, we start with the left-right symmetric 
model shown in Table 4.1. 
\bigno
\cl{\vbox{
\hbox{\vbox{\offinterlineskip
\def\tablespace{height2pt&\omit&&\omit&&\omit&\cr}
\def\tablerule{\tablespace\noalign{\hrule}\tablespace}
\def\Tablerule{\tablespace\noalign{\hrule height1pt}\tablespace}
\hrule\halign{&\vrule#&\strut\hskip0.2cm\hfil#\hfill\hskip0.2cm\cr
\tablespace
& part && $c$ && $\o{c}$ &\cr
\tablerule\tablerule
& $4D$ space-time, $X^{\mu}$ && $2$ && $2$ & \cr
\tablerule
& $N=2$ Virasoro && $9$ && $9$ &\cr
\tablerule
& $U(1)_2$ && $1$ && $1$ &\cr
\tablerule
& gauge group\ $SO(8)$ && $4$ && $4$ &\cr
\Tablerule
& $U(1)_2\times U(1)_2$ && $2$ && $2$ &\cr
\tablerule
& gauge group\ $SO(12)$ && $6$ && $6$ &\cr
\tablespace}\hrule}}
\hbox{\hskip 0.5cm Table 4.1 \hskip 0.5cm Underlying CFT for 
 $SO(10)\times E_7$}}}
\meno
The whole partition function is now written in terms of simple current 
modular invariants in the following way:
$$\eqalignno{Z\sim\vec{\chi}(\tau)\,&{\cal M}({\cal J}_{SO(12)\to E_7})\,
           M(J_{GSO_L})\,\prod_l M(\Upsilon_l)\,\,M(J_{GSO_R})\,&(4.1)\cr
           &\,\prod_i M(J_i)\,M(J_{ SO(8)\to SO(10)})\,
           \prod_{k=1}^2{\cal M}({\cal J}^k_{SO(12)\to E_8})\,\,
           \vec{\chi}(\o\tau) &\cr} $$
where the simple currents ${\cal J}$ are those new ones 
which only act on the $(U(1)_2)^2\times SO(12)$ piece\footnote{$^2$}
{For the meaning of our notation and the explicit form of the simple 
currents $J$ please take a look into [\selfa].}:
$$\eqalignno{ {\cal J}_{SO(12)\to E_7}&=\phi^{U(1)_2}_{1,2}\otimes
             \phi^{U(1)_2}_{1,2}\otimes\phi^{SO(12)}_s &\cr
            {\cal J}^1_{SO(12)\to E_8}&=\phi^{U(1)_2}_{2,2}\otimes
             \phi^{U(1)_2}_{2,2}\otimes \phi^{SO(12)}_0 &(4.2)\cr
            {\cal J}^2_{SO(12)\to E_8}&=\phi^{U(1)_2}_{1,2}\otimes
             \phi^{U(1)_2}_{1,2}\otimes\phi^{SO(12)}_s. &\cr }$$
The last two simple currents guarantee that the right moving side is 
extended to $E_8$ so that the bosonic string map can still be applied to 
obtain at the end of the day a heterotic string. Unlike those models with 
a hidden $E_8$ gauge symmetry, for these models the left-right symmetry 
breaking simple currents $\Upsilon_l$ are also allowed to contain 
contributions from the $(U(1)_2)^2\times SO(12)$ sector.
\vfill\eject\pano
\section{5.\ Examples }
\meno
Now, we are searching for a parent tensor product of $N=2$ minimal 
models and a set of simple currents $\Upsilon_l$ which both yield an 
$N=2$ space-time supersymmetric spectrum and break the hidden $E_8$ down 
to $E_7$. Fortunately, after some failing attempts our computer program 
provides us with the desired models.
\meno
\item {a.)} We use the $(1)(4)^4$ parent model and choose a certain set of 
three further $\Upsilon_l$'s yielding the following massless spectrum with 
gauge group $E_7\times E_7\times U(1)_T^2\times SU(3)_T\times G(10)$ where 
only $SU(3)_T$ singlets appear in the spectrum:\ the massless sector is
$$ N_{56,1}=4,\quad N_{1,56}=4,\quad N_{1,1}=72 .\eqno(5.1)$$ 
The geometric interpretation is as follows:\ compactify the 
ten-dimensional string on a torus down to eight dimensions yielding an 
enhanced $E_8\times E_8\times SU(3)_T\times U(1)^2_T$ gauge group. Then 
embed an $SU(2)$ bundle with $A^{E_8}_2=12$ into each of the two $E_8$ 
factors and let the toroidal $SU(3)_T$ unbroken. The massless spectrum is 
given by the index theorem (2.1):
$$ N_{56,1}=4,\quad N_{1,56}=4,\quad N_{1,1}=62.\eqno(5.2)$$
After higgsing away the $G(10)$ piece we obtain complete agreement of the 
two spectra.
\bigno
\item {b.)} We find another example of this kind with gauge group
$E_7\times E_7\times U(1)_T^2\times G(14)$:
$$ N_{56,1}=2,\quad N_{1,56}=2,\quad N_{1,1}=76.\eqno(5.3)$$
Here, we embed an $SU(2)$ bundle with $A^{E_8}_2=8$ into each of the two 
$E_8$ factors and an $SU(3)$ bundle with $A^{SU(3)}_3=8$ into the toroidal
$SU(3)_T$. The massless spectrum is 
$$ N_{56,1}=2,\quad N_{1,56}=2,\quad N_{1,1}=62.\eqno(5.4)$$
After higgsing completely the enhanced gauge group $G(14)$ of our model we 
get exactly this spectrum.
\bigno
\item{c.)} The next example has gauge group 
$SO(10)\times E_7\times U(1)_T^2\times G(10)$ and
$$ N_{16,1}=8,\quad N_{10,1}=10,\quad N_{1,1}=88.\eqno(5.5)$$
After higgsing the $G(10)$ the spectrum agrees with the one obtained by 
embedding an $SU(4)$ bundle with $A^{E_8}_4=16$ into the first $E_8$, an 
$SU(2)$ bundle with $A^{E_8}_2=4$ into the second $E_8$ and an $SU(3)$ 
bundle with $A^{SU(3)}_3=4$ into the toroidal $SU(3)_T$.
\bigno
\item{d.)} We also obtain a slightly different example with gauge group 
$SO(10)\times E_7\times U(1)_T^2\times G(8)$ and
$$N_{16,1}=4,\quad N_{10,1}=6,\quad N_{1,1}=80,\quad N_{1,56}=1.
 \eqno(5.6)$$
After higgsing the $G(8)$ the spectrum agrees with the one obtained by 
embedding an $SU(4)$ bundle with $A^{E_8}_4=12$ into the first $E_8$, an 
$SU(2)$ bundle with $A^{E_8}_2=6$ into the second $E_8$ and an $SU(3)$ 
bundle with $A^{SU(3)}_3=6$ into the toroidal $SU(3)_T$.
\meno
There are many other examples which all have an interpretation in terms
of embeddings of suitable bundles into both $E_8$ factors. Thus, we have 
succeeded in formulating the SCFT analogue of choosing arbitrary bundles 
on $K_3$. Clearly, there is no restriction to consider only $N=2$ models 
so that we claim to have an SCFT analogue of embedding vector bundles on 
general Calabi-Yau manifolds into both $E_8$ factors. This is the general 
class of $(0,2)$ supersymmetric (non)linear $\sigma$-models studied 
intensively by J.\ Distler and S.\ Kachru in [\diska]. 
Furthermore, the generalization to embeddings of higher rank bundles into 
the second $E_8$ is obvious, one starts with $(U(1)_2)^r\times SO(16-2r)$ 
in the former $E_8$ sector and introduces suitable simple currents to 
extend the sector on the right moving side to $E_8$ and on the left moving 
side to $E_{9-r}$\footnote{$^3$}{We define $E_5=SO(10)$ and $E_4=SU(5)$.}. 
\bigno
In the following we will investigate example a.) in more detail showing
that it gives rise to a lot of possible heterotic/\type dual pairs. By 
looking at the massless sector we find that actually the fully enhanced 
gauge group is
$$G=E_7\times E_7\times SU(3)_E\times U(1)_E^2
 \times SU(3)_T\times U(1)^2_T\eqno(5.7)$$
where the subscript $T$ indicates the gauge factors arising from the torus 
and the subscript $E$ the special enhanced gauge symmetry. The massless 
sector fits into the following representations: 
$$4\times(56,1,1)+4\times(1,56,1)+18\times(1,1,3)+18\times(1,1,1).
 \eqno(5.8)$$
Furthermore, there are singlet fields which carry nonzero charge with 
respect to the two $U(1)_E$ currents. Of course, all states are singlets 
with respect to the toroidal gauge group. Besides the special enhanced 
gauge symmetry, this model has already been mentioned in the original 
paper of S.\ Kachru and C.\ Vafa [\kachru] where it was argued that by 
completely higgsing away the two $E_7$ factors and the 
$SU(3)_E\times U(1)^2_E$ factor one yields a spectrum 
$(n_H,n_V)=(244,4)$. The hypersurface $\P_{1,1,2,8,12}[24]$ has Hodge 
numbers $(b_{21},b_{11})=(243,3)$ being a good candidate for the \typea
dual model. We find further candidates for heterotic/\typea dual 
pairs in the moduli space of this model. 
\pano
In Table 5.1 we describe all possible dual pairs which we find in the 
moduli space of this model by successively breaking parts of the gauge 
group $G$. The second column shows to which subgroup the gauge group is 
broken before switching on VEVs for the remaining generators in the u
adjoint representations of the nonabelian sector.
\bigno
\cl{\vbox{
\hbox{\vbox{\offinterlineskip
\def\tablespace{height2pt&\omit&&\omit&&\omit&\cr}
\def\tablerule{\tablespace\noalign{\hrule}\tablespace}
\hrule\halign{&\vrule#&\strut\hskip0.2cm\hfil#\hfill\hskip0.2cm\cr
\tablespace
& $(n_H,n_V)$ && $G'\subset E_7\times E_7\times SU(3)_E\times U(1)_E^2
 \times SU(3)_T\times U(1)^2_T$ && $K_3$ fibration &\cr
\tablerule\tablerule
& $(244,4)$ && $U(1)_T^4$ && $\P_{1,1,2,8,12}[24]$ & \cr
\tablespace
& $(165,9)$ && $SO(10)\times U(1)_T^4$ && $\P_{1,1,4,6,12}[24]$ & \cr
\tablespace
& $(128,8)$ && $SO(4)\times SO(4)\times U(1)_T^4$ && 
 $\P_{2,3,3,16,24}[48]$ & \cr
\tablespace
& $(97,13)$ && $SO(10)\times SO(8)\times U(1)_T^4$ && 
 $\P_{1,1,6,8,8,10}[18\, 16]$ & \cr
\tablerule
& $(123,11)$ && $SU(4)\times SU(4)\times U(1)_E\times U(1)_T^4$ && 
 $\P_{1,1,4,8,10,12}[20\,16]$ & \cr
\tablespace
& $(108,12)$ && $SU(5)\times SU(4)\times U(1)_E\times U(1)_T^4$ && 
 $\P_{3,3,4,20,30}[60]$ & \cr
\tablespace
& $(105,13)$ && $SO(10)\times SU(4)\times U(1)_E\times U(1)_T^4$ && 
 $\P_{1,1,6,8,10,10}[20\, 16]$ & \cr
\tablerule
& $(132,12)$ && $E_6 \times SU(2)_E\times U(1)_E\times U(1)_T^4$ && 
 $\P_{1,1,6,8,8}[24]$ & \cr
\tablespace
& $(126,10)$ && $SU(4)\times SU(2)\times SU(2)_E\times U(1)_E
 \times U(1)_T^4$ && $\P_{1,1,4,6,8}[20]$ & \cr
\tablespace
& $(109,13)$ && $SO(10)\times SU(2)\times SU(2)_E\times U(1)^2_E
 \times U(1)_T^4$ && $\P_{1,1,6,8,10,12}[20\, 18]$ & \cr
\tablerule
& $(154,10)$ && $SU(3) \times SU(3)_E\times U(1)^2_E\times U(1)_T^4$ && 
 $\P_{1,1,4,8,10}[24]$ & \cr
\tablespace
& $(112,10)$ && $SO(4)\times SU(2)\times SU(3)_E\times U(1)_E
 \times U(1)_T^4$ && $\P_{1,1,4,6,6}[18]$ & \cr
\tablespace
& $(82,10)$ && $SO(4)\times SO(4)\times SU(3)_E\times U(1)_T^4$ && 
 $\P_{1,1,4,4,6,8}[12\,12]$ & \cr
\tablespace
& $(63,15)$ && $SU(5)\times SU(4)\times SU(3)_E\times U(1)^2_E
 \times U(1)_T^4$ && $\P_{3,3,8,10,24}[48]$ & \cr
\tablespace
& $(40,16)$ && $SO(10)\times SO(10)\times SU(3)_E\times U(1)_T^4$ && 
 $\P_{2,4,4,5,5}[20]$ & \cr
\tablespace}\hrule}}
\hbox{\hskip 0.5cm Table 5.1 \hskip 0.5cm Heterotic/\typea duals }}}
\meno
Note that only the first four dual pairs could be found by knowing only
the generic gauge group, the entire rest has a nontrivial contribution
from the special $SU(3)_E\times U(1)^2_E$ enhanced gauge symmetry. 
So far we have only dealt with heterotic/\typea duality. However, by 
looking into the list of [\lyn] we find also some possible 
heterotic/\typeb dual pairs that are presented in Table 5.2. For the 
\typeb string compactified onto a Calabi-Yau threefold the $N=2$ massless
spectrum is $(n_H,n_V)=(b_{11}+1,b_{21}+1)$. 
\bigno
\cl{\vbox{
\hbox{\vbox{\offinterlineskip
\def\tablespace{height2pt&\omit&&\omit&&\omit&\cr}
\def\tablerule{\tablespace\noalign{\hrule}\tablespace}
\hrule\halign{&\vrule#&\strut\hskip0.2cm\hfil#\hfill\hskip0.2cm\cr
\tablespace
& $(n_H,n_V)$ && $G'\subset E_7\times E_7\times SU(3)_E\times U(1)_E^2
 \times SU(3)_T\times U(1)^2_T$ && $K_3$ fibration &\cr
\tablerule\tablerule
& $(42,18)$ && $SO(10)\times SO(10)\times SU(3)_E\times U(1)_E^2
 \times U(1)_T^4$ && $\P_{4,9,9,10,22}[54]$ & \cr
\tablespace
& $(35,17)$ && $E_6\times SO(10)\times SU(3)_E\times U(1)_T^4$ && 
 $\P_{4,8,10,11,11}[44]$ & \cr
\tablespace
& $(30,18)$ && $E_6\times E_6\times SU(3)_E\times U(1)_T^4$ && 
 $\P_{8,10,12,15,15}[60]$ & \cr
\tablespace}\hrule}}
\hbox{\hskip 0.5cm Table 5.2 \hskip 0.5cm Heterotic/\typeb duals }}}
\bigno
\bigno
By looking more closely at example b.) we find that the complete gauge 
group is
$$ G=E_7\times E_7\times SU(3)_E\times SU(2)_E\times U(1)_E^3 
 \times U(1)^2_T.\eqno(5.9)$$ 
The massless spectrum can be reduced out with respect to this gauge group:
$$\eqalignno{&2\times(56,1,1,1)+2\times(1,56,1,1)+4\times(1,1,3,2)+&\cr
 &10\times(1,1,3,1)+4\times(1,1,1,2)+14\times(1,1,1,1).&(5.10)\cr}$$
Despite the fact that the two $E_7$ factors minimally can be broken down 
only to $SO(8)$ we find some further possible heterotic/\typea dual 
pairs in the different Higgs phases of this model given in Table 5.3. 
\bigno
\cl{\vbox{
\hbox{\vbox{\offinterlineskip
\def\tablespace{height2pt&\omit&&\omit&&\omit&\cr}
\def\tablerule{\tablespace\noalign{\hrule}\tablespace}
\hrule\halign{&\vrule#&\strut\hskip0.2cm\hfil#\hfill\hskip0.2cm\cr
\tablespace
& $(n_H,n_V)$ && $G'\subset E_7\times E_7\times SU(3)_E\times SU(2)_E
 \times U(1)_E^3 \times U(1)^2_T$ && $K_3$ fibration &\cr
\tablerule\tablerule
& $(76,10)$ && $SO(8)\times SO(8)\times U(1)_T^2$ && 
 $\P_{1,1,4,4,6,6}[12\, 10]$ & \cr
\tablespace
& $(77,11)$ && $SO(8)\times SO(8)\times U(1)_E\times U(1)_T^2$ && 
 $\P_{1,1,4,6,6,6}[12\, 12]$ & \cr
\tablespace
& $(47,11)$ && $SO(8)\times SO(8)\times SU(2)_E\times U(1)_T^2$ && 
 $\P_{4,5,5,6,20}[40]$ & \cr
\tablespace
& $(44,12)$ && $SO(10)\times SO(8)\times SU(2)_E\times U(1)_T^2$ && 
 $\P_{6,7,7,8,28}[56]$ & \cr
\tablespace
& $(40,16)$ && $E_6\times E_6\times SU(2)_E\times U(1)_E\times U(1)_T^2$ &&
 $\P_{2,4,4,5,5}[20]$ & \cr
\tablespace}\hrule}}
\hbox{\hskip 0.5cm Table 5.3 \hskip 0.5cm Heterotic/\typea duals }}}
\meno
Only the last model was already exhibited in Table 5.1. Furthermore, there 
exist two heterotic/ \typeb dual pairs contained in Table 5.4.
\bigno
\cl{\vbox{
\hbox{\vbox{\offinterlineskip
\def\tablespace{height2pt&\omit&&\omit&&\omit&\cr}
\def\tablerule{\tablespace\noalign{\hrule}\tablespace}
\hrule\halign{&\vrule#&\strut\hskip0.2cm\hfil#\hfill\hskip0.2cm\cr
\tablespace
& $(n_H,n_V)$ && $G'\subset E_7\times E_7\times SU(3)_E\times SU(2)_E
 \times U(1)_E^3 \times U(1)^2_T$ && $K_3$ fibration &\cr
\tablerule\tablerule
& $(45,15)$ && $E_6\times SO(8)\times SU(2)_E\times U(1)^2_E
 \times U(1)_T^2$ && $\P_{5,5,12,16,22}[60]$ & \cr
\tablespace
& $(34,18)$ && $E_7\times E_7\times SU(2)_E\times U(1)_E\times U(1)_T^2$ &&
 $\P_{5,5,6,8,16}[40]$ & \cr
\tablespace}\hrule}}
\hbox{\hskip 0.5cm Table 5.4 \hskip 0.5cm Heterotic/\typeb duals }}}
\meno
Thus, by studying the enhanced gauge symmetries of only four exactly 
solvable $N=2$ models we found already 32 different possible 
heterotic/\typea and 5 heterotic/\typeb dual pairs. Since the known lists 
of $K_3$ fibrations are likely far from being complete there might exist 
even more heterotic strings in the Higgs phases of these models which admit
a \type dual description.
\bigno
\section{6.\ Conclusions }
\meno
We have studied Gepner-like exactly solvable models with $(0,4)$ world 
sheet supersymmetry leading to $N=2$ space-time supersymmetric spectra. 
Unlike the more complicated case with $(0,2)$ world sheet supersymmetry, 
these models all fit into the scheme of nonlinear $\sigma$-models on 
$K_3\times T_2$ with nontrivial embeddings of vector bundles into the 
gauge degrees of freedom. 
\pano
In addition, we have extended the class of Gepner-like $(0,2)$ SCFTs to the
case where also the hidden $E_8$ gauge group is broken down to some 
subgroup.
\pano
Furthermore, by calculating the complete enhanced gauge symmetry of a few 
examples we have found interesting branches of the heterotic $N=2$ moduli 
space leading to some candidates of heterotic/\type dual models. 
\meno
{\bf Acknowledgements}
\smno
It is a pleasure to thank L.\ Dolan, S.\ Kachru and W.\ Nahm for 
discussion. This work is supported by U.S.\ DOE grant No.\ 
DE-FG05-85ER-40219.
\baselineskip=10.3pt
\bigno
\section{References}
\meno
\bibitem{\alda} G.\ Aldazabal, A.\ Font, L.\ Ib\'a$\tilde{\rm n}$ez and
F.\ Quevedo, {\it Chains of $N=2$, $D=4$ heterotic type$\,II$ duals}, 
hep-th/9510093
\bibitem{\antone} I.\ Antoniadis, E.\ Gava, K.S.\ Narain and T.R.\ Taylor,
{\it $N=2$ type$\,II$ heterotic duality and higher derivative F-terms},
Nucl.\ Phys.\ {\bf B455} (1995) 109, hep-th/9507115
\bibitem{\antonz} I.\ Antoniadis and H.\ Partouche, {\it Exact monodromy 
group of $N=2$ heterotic superstring}, preprint CPTH-RR370.0895, 
hep-th/9509009
\bibitem{\aspina} P.\ Aspinwall and J.\ Louis, {\it On the ubiquity of 
$K_3$ fibrations in string duality}, hep-th/9510234
\bibitem{\aspinb} P.\ Aspinwall, {\it Enhanced gauge symmetries and 
Calabi-Yau threefolds}, hep-th/9511171
\bibitem{\selfa} R.\ Blumenhagen and A.\ Wi{\ss}kirchen, {\it Exactly 
solvable $(0,2)$ supersymmetric string vacua with GUT gauge groups},
Nucl.\ Phys.\ {\bf B454} (1995) 561, hep-th/9506104 
\bibitem{\selfb} R.\ Blumenhagen, A.\ Wi{\ss}kirchen and R.\ Schimmrigk,
{\it The $(0,2)$ exactly solvable structure of chiral rings, 
Landau-Ginzburg theories and Calabi-Yau manifolds}, preprint IFP-601-UNC, 
BONN-TH-95-17, NSF-ITP-95-120, hep-th/9510055, to be published in 
Nucl.\ Phys.\ {\bf B}
\bibitem{\diska} J.\ Distler and S.\ Kachru, {\it $(0,2)$ Landau-Ginzburg
theory}, Nucl.\ Phys.\ {\bf B413} (1994) 213, hep-th/9309110
\bibitem{\ferrara} S.\ Ferrara, J.\ Harvey, A.\ Strominger and C.\ Vafa,
{\it Second quantized mirror symmetry}, Phys.\ Lett.\ {\bf B361} (1995) 59,
hep-th/9505162
\bibitem{\gepe} D.\ Gepner, {\it Space-time supersymmetry in compactified 
string theory and superconformal models}, Nucl.\ Phys.\ {\bf B296} (1988) 
757
\bibitem{\greene} B.\ Greene, D.\ Morrison and A.\ Strominger,
{\it Black hole condensation and the unification of string vacua},
Nucl.\ Phys.\ {\bf B451} (1995) 109, hep-th/9504145
\bibitem{\intri} K.\ Intriligator, {\it Bonus symmetry in conformal
field theory}, Nucl.\ Phys.\ {\bf B332} (1990) 541
\bibitem{\kachru} S.\ Kachru and C.\ Vafa, {\it Exact results for $N=2$ 
compactifications of heterotic strings}, Nucl.\ Phys.\ {\bf B450} (1995) 
69, hep-th/9505105
\bibitem{\kakl} S.\ Kachru, A.\ Klemm, W.\ Lerche, P.\ Mayr and C.\ Vafa,
{\it Nonpertubative results on the point particle limit of $N=2$ 
heterotic string compactifications}, hep-th/9508155
\bibitem{\kaplu} V.\ Kaplunovsky, J.\ Louis and S.\ Theisen, {\it Aspects 
of duality in $N=2$ string vacua}, Phys.\ Lett.\ {\bf B357} (1995) 71, 
hep-th/9506110
\bibitem{\klemm} A.\ Klemm, W.\ Lerche and P.\ Mayr, {\it $K_3$ fibrations 
and heterotic type$\,II$ string duality}, Phys.\ Lett.\ {\bf B357} (1995) 
313, hep-th/9506112
\bibitem{\lyn} M.\ Lynker and R.\ Schimmrigk, {\it Conifold transitions 
and mirror symmetries}, hep-th/9511058
\bibitem{\sche} A.N.\ Schellekens and S.\ Yankielowicz, {\it Extended 
chiral algebras and modular invariant partition functions}, Nucl.\ Phys.\
{\bf B327} (1989) 673
\bibitem{\schz} A.N.\ Schellekens and S.\ Yankielowicz, {\it Modular 
invariants from simple currents. An explicit proof}, Phys.\ Lett.\ 
{\bf B227} (1989) 387
\bibitem{\schd} A.N.\ Schellekens and S.\ Yankielowicz, {\it New modular 
invariants for $N=2$ tensor products and four-dimensional strings}, 
Nucl.\ Phys.\ {\bf B330} (1990) 103
\bibitem{\schv} A.N.\ Schellekens and S.\ Yankielowicz, {\it Simple 
currents, modular invariants and fixed points}, Int.\ J.\ Mod.\ Phys.\ 
{\bf A5} (1990) 2903
\bibitem{\sei} N.\ Seiberg and E.\ Witten, {\it Electric-magnetic duality,
monopole condensation and confinement}, Nucl.\ Phys.\ {\bf B426} (1994) 19,
hep-th/9407087; Erratum, Nucl.\ Phys.\ {\bf B430} (1994) 485
\bibitem{\stro} A.\ Strominger, {\it Massless black holes and conifolds in
string theory}, Nucl.\ Phys.\ {\bf B451} (1995) 96, hep-th/9504090
\bibitem{\vafa} C.\ Vafa and E.\ Witten, {\it Dual string pairs with $N=1$
and $N=2$ supersymmetry in four dimensions}, preprint HUTP-95/A023, 
IASSNS-HEP-95-58, hep-th/9507050
\vfill\end